\documentclass[prd,aps,nofootinbib,notitlepage]{revtex4}%showpacs,showkeys,preprintnumbers
\usepackage{graphicx,epsf,amsmath,amsfonts,amssymb,amsbsy}
\usepackage{epsfig}
\usepackage{epstopdf}

\newcommand{\ds}{\displaystyle}
\newcommand{\vev}[1]{\langle#1\rangle}
\newcommand{\mat}{\left ( \begin{array}}
\newcommand{\emat}{\end{array} \right )}
\newcommand{\vect}{\left ( \begin{array}{c}}
\newcommand{\evect}{\end{array} \right )}

\newcommand{\Det}{\mathop{\rm Det}\nolimits}
\newcommand{\e}{\mathop{\rm e}\nolimits}
\begin{document}

\title{ \bf  Dualities and inhomogeneous phases in dense quark matter with chiral and isospin imbalances in the framework of effective model
}
\author{T. G. Khunjua$^{1}$,  K. G. Klimenko$^{2}$, R. N. Zhokhov$^{3}$}
\vspace{1cm}
\affiliation{$^{1)}$ Faculty of Physics, Moscow State University,
119991, Moscow, Russia}
 \affiliation{$^{2)}$ Logunov Institute for High Energy Physics,
NRC "Kurchatov Institute", 142281, Protvino, Moscow Region, Russia}
\affiliation{$^{3)}$  Pushkov Institute of Terrestrial Magnetism, Ionosphere and Radiowave Propagation (IZMIRAN),
108840 Troitsk, Moscow, Russia}

\begin{abstract}
It has been shown in \cite{kkz3,kkz2} in the framework of Nambu--Jona-Lasinio model with the assumption of spatially homogeneous 
condensates that in the large-$N_{c}$ limit ($N_{c}$ is the number of quark colours) there exist three dual symmetries of the 
thermodynamic potential, which describes dense quark matter with chiral and isospin imbalances.
The main duality is between the chiral symmetry breaking and the charged pion condensation phenomena. There have been a lot of 
studies and hints that the ground state could be characterized by spatially inhomogeneous condensates, so the question arises if 
duality is a rather deep property of the phase structure or just accidental property in the homogeneous case. In this paper we 
have shown that even if the phase diagram contains phases with spatially inhomogeneous condensates, it still possesses the 
property of this main duality. Two other dual symmetries are not realized in the theory if it is investigated within an 
inhomogeneous approach to a ground state. Based on various previously studied aspects of the QCD phase diagram of dense isospin 
asymmetric matter with possible inhomogeneous condensates, in the present paper a unified picture and full phase diagram of dense 
quark matter with isospin imbalance have been assembled. %From various earlier studied aspects of QCD phase diagram of isospin asymmetric matter 
%with possible inhomogeneous condensates, in the present paper the unified picture and full phase diagram of isospin imbalanced 
%dense quark matter have been assembled. 
Acting on this diagram by a dual transformation, we obtained, in the framework of an approach with spatially inhomogeneous 
condensates and without any calculations, a full phase diagram of chirally asymmetric dense medium. This example shows that the 
duality is not just entertaining mathematical property but an instrument with very high predictivity power. The obtained phase 
diagram is quite rich and contains various spatially inhomogeneous phases.
\end{abstract}

%\pacs{11.30.Qc, 12.39.-x, 21.65.+f}
% 11.30.Qc Spontaneous and radiative symmetry breaking
% 12.39.-x Phenomenological quark models
% 11.15.Ex Spontaneous breaking of gauge symmetries
% 12.38.Aw General properties of QCD
% 12.38.Mh Quark-gluon plasma
% 11.10.St Bound and unstable states; Bethe-Salpeter equations
% 12.38.-t Quantum chromodynamics
% 12.38.Lg Other nonperturbative calculations
% 26.60.+c Nuclear matter aspects of neutron stars
% 21.65.+f Nuclear matter
% 12.39.Fe

%\keywords{Nambu -- Jona-Lasinio model; Color superconductivity;
%Nambu--Goldstone bosons
%}
\maketitle
%\draft
%\large
%\maketitle

\section{Introduction}

The phase diagram of strongly interacting matter (QCD phase diagram) 
is now one of the open questions in the Standard Model of elementary 
particle physics. The knowledge of QCD phase structure is essential for the understanding of such physical phenomena as compact stars  
and laboratory experiments on relativistic heavy-ion collisions.
While the properties of matter at finite temperatures and vanishing densities are now understood quite well thanks to 
first principle lattice calculations and heavy-ion collisions, there is no consensus yet on the QCD phase structure at finite chemical potentials. In this case lattice simulations are 
not possible up to now and most of our understanding comes from calculations performed within effective models. One of the most used models is Nambu--Jona-Lasinio (NJL) model \cite{Nambu:1961fr} (see for review, e.g., in Refs. \cite{Hatsuda:1994pi}).

Besides the temperature and baryon chemical potential $\mu_B$ it is interesting to take into account nonzero isotopic $\mu_I$ 
and chiral chemical potentials. Isotopic (isospin) chemical potential allows us to consider systems with isospin imbalance (different numbers of $u$ and $d$ quarks). %There is another enthralling phenomenon,  
Chiral chemical potential $\mu_{5}$ accounts for a chiral imbalance (or density) $n_{5}$ (it is the difference between densities of 
right-handed and left-handed quarks). It has been argued that the chiral imbalance $n_5$ (or $\mu_{5}$) can be
generated dynamically at high temperatures, for example, in the fireball after heavy ion collision, by virtue of the Adler-Bell-Jackiw 
anomaly and quarks interacting with gauge (gluon) field configurations with nontrivial topology, named sphalerons.  
On par with an external strong magnetic field, which can be produced in heavy ion collisions as well, this can lead to 
the so-called chiral magnetic effect \cite{fukus}. Moreover, in the presence of external magnetic field chiral density 
$n_5$ can be produced in dense quark matter due to the so-called chiral separation effect \cite{Metlitski}.  Similar in essence, 
the generation mechanism of $n_5\ne 0$ can be envisaged also in fast rotating dense matter due to the so-called chiral 
vortical effect \cite{Fukushima:2018grm}. Chiral imbalance can
also be considered as a phenomenon generated within the framework of pure quantum electrodynamics, by coupling quarks to external 
parallel electric and magnetic fields \cite{Ruggieri:2016fny, Ruggieri:2016fny1}. This mechanism, in our reckoning, is very 
promising and can lead to a rather large chiral imbalance in the system.

Now let us notice that usually when one talks about chiral density $n_5$ (or $\mu_{5}$) one implies that separately chiral 
densities of $u$ and $d$ quarks are equal to each other. Indeed, that is the case when one has in mind the mechanism of generation of 
chiral imbalance at high temperatures due to nontrivial gauge field configurations. It is obvious due to the fact that gluon 
field interact with different quark flavours in exactly the same way and does not feel the difference between flavours.
But other mechanisms may be sensitive to the flavour of quarks and we will now elaborate on that. The concept of chiral density 
(imbalance) can also be introduced for each flavour of quarks. So from now on $n_{5u}$ and  $n_{5d}$ are the notations for chiral 
imbalance of $u$ and $d$ quarks, respectively. It is evident that $n_5=n_{5u}+n_{5d}$ and, correspondingly, 
$\mu_5=\mu_{5u}+\mu_{5d}$. But there is a quantity that has not been considered above, namely $n_{I5}=n_{5u}-n_{5d}$ (and, 
correspondingly, $\mu_{I5}=\mu_{5u}-\mu_{5d}$), which is called chiral isospin density (imbalance).

As it has been already noted, the mechanism for generation of the chiral imbalance at high temperatures due to  
nontrivial gauge field configurations does not work for chiral isospin imbalance, and in this case we have $n_{I5}=0$. 
But other mechanisms that were discussed above can lead to its generation, $n_{I5}\ne 0$. 
It can be shown that chiral isospin density $n_{I5}$ can be created in a line with chiral one in dense magnetized quark matter 
(due to the chiral separation effect) (see in Appendix A of the present paper) or in rotating dense quark matter 
(due to the chiral vortical effect). Moreover, due to the different quark electric charges, in parallel 
electric and magnetic fields one can get equally successful both chiral and chiral isospin imbalances. 
This mechanism provides a rather considerable chiral imbalance, both with $n_5\ne 0$ and $n_{I5}\ne 0$, 
in quark systems, especially in heavy-ion collisions when very strong electric and magnetic fields can be produced in the 
direction both parallel and perpendicular to the reaction plane \cite{Bzdak:2011yy}.
It probably could be viable scenario also in neutron stars.

Even though there are several mechanisms of generating chiral isospin imbalance $n_{I5}\ne 0$, 
they have not been studied very comprehensively yet and the consideration of chiral isospin imbalance can look academic. 
But even if one imagines that in physical settings chiral isospin imbalance is absent, still, it could be interesting to study 
its influence on phase structure of quark matter. Both situations, $\mu_{5}\neq0$ and $\mu_{I5}\neq0$, describe chiral imbalance 
in the system. If one investigates an influence of chiral imbalance, in general, on some phenomenon for the first time 
(inhomogeneous phases have not been studied with chiral imbalance of any form) and wants to get some insight, one can first 
study any of the two types of chiral imbalances that is easier to investigate. In our case one can study the influence of the 
$n_{I5}\ne 0$ ($\mu_{I5}\ne 0$) on the inhomogeneous phases just by duality notion (will be discussed below) without performing 
any numerical calculation (and they are rather involved). This example also nicely demonstrates the strength and the beauty of 
the dulity. Of course, since both $\mu_{I5}$ and $\mu_{5}$ describes chiral imbalance in the system, one can anticipate that 
the influence of both these imbalances can have similar, at least general, features. For example, both $\mu_{I5}$ and $\mu_{5}$ 
generate charged pion condensation (CPC) phase in dense quark matter, although, this generation is various in 
intensity, or in homogeneous case their influences on chiral symmetry breaking (CSB) are identical at zero baryon and isospin 
densities \cite{kkz2}. Let us also recall that sometimes it is possible that consideration of unphysical settings can lead to 
considerable advances in understanding of real physical phenomena. For example, in Ref. \cite{Son:2000xc} large isospin density 
(relative to baryon one) was considered, and it has taught us a great deal about quark pairing and the properties of dense matter. 

Moreover, in the context of QCD phase diagram the inclusion of $\mu_{I5}$ is formally more rigorous, than the consideration of 
$\mu_{5}$.
Indeed, the consideration of QCD with chiral imbalance, when $\mu_{5}\ne 0$, does have an internal problems, 
$\mu_{5}$ is not conjugated to a quantity that is strictly conserved in strong interacting matter due to already mentioned 
Adler-Bell-Jackiw quantum anomaly and chirality changing processes. It can be easily seen in the framework of the NJL model, 
chiral charge $n_{5}$ is conjugated to $U_{A}(1)$ symmetry
(which is not conserved due to quantum anomaly) and one can easily see that this symmetry is explicitly broken by NJL model 
interaction. So the introduction of $\mu_{5}$ is troublesome, although it is massively done in the literature. 
It is pointed out, however, in Ref. \cite{Ruggieri:2016fny1} that it is possible to consider $\mu_{5}$, if one assumes that the 
system is observed on a time scale much larger than the typical time scale
of the chirality changing processes. Let us note that there is no such problem with chiral isospin imbalance, i.e. when 
$\mu_{I5}\ne 0$. In this case in the chiral limit  $U_{IA}(1)$ symmetry (chiral isospin charge $n_{I5}$ is conjugated to) is 
anomaly free and conserved in the NJL model.

Recently, it was found a number of strong arguments, corroborated by model calculations, %which have 
implying that at large and intermediate densities the phases with spatially inhomogeneous order parameters (condensates) can exist in different branches of modern physics. If ground state of a theory is inhomogeneous, then some of its spatial symmetries are spontaneously broken. This phenomenon is not a quite new concept. 
For example, the idea of density waves in nuclear matter was already discussed in 1960 by Overhauser \cite{Overhauser1960}. Moreover, in the 1970s there was much activity related to inhomogeneous pion condensation \cite{BrownWeise}, first considered by Migdal \cite{Migdal}. Spatially inhomogeneous phases in nucleon matter were also studied in \cite{Dautry:1979bk} in terms of sigma model for nucleon and meson fields. These phases in the form of
charge and spin density waves are not exotic in condensed matter physics as well \cite{Gruner}. In particular, the crystalline
phases have been considered long time ago for superconductors by Larkin and Ovchinikov \cite{LarkinOvchinnikov} as well as by Fulde and Ferrell \cite{FuldeFerrell} (usually, the phase is called LOFF phase), etc.
And more recently crystalline inhomogeneous phases in color superconducting quark matter was discussed in Refs. \cite{Alford:2000ze,%} %\cite{
Anglani:2013gfu, Casalbuoni:2003wh, Bowers:2002xr, Mannarelli:2015bda}.
Deryagin, Grigoriev, and Rubakov (DGR) have shown that at high densities in the limit of infinite number of colors $N_{c}$ the QCD ground state might be inhomogeneous and anisotropic so that the QCD ground state has the structure of a spatial standing wave \cite{dgr}. And of course 
spatially nonuniform condensates are intensively studied in the framework of effective theories such as NJL$_4$ model \cite{Sadzikowski:2000ap} or different (1+1)-dimensional theories with four-fermionic interactions \cite{Basar:2009fg}. The modern state of investigations of dense baryonic matter in the framework of the spatially inhomogeneous condensate approach is presented in the
reviews \cite{buballa,Heinz:2014pkl}.

Unfortunately, general analytical methods, which can be used in order to establish the inhomogeneity of the ground state of an arbitrary system, are absent. So different ansatzes for spatially inhomogeneous condensates are used. For example, for charged pion and/or superconducting condensates usually the single-plane-wave LOFF ansatz is employed \cite{he,ebert}. The simplest possible spatially inhomogeneous ansatz for chiral condensate in quantum field theories is an analog of the spin-density waves in condensed matter systems, and it is called ``chiral density wave`` (CDW) \cite{nakano}. Being an analytically treatable case, CDW ansatz has been the object of intense investigations during the course of the last 25 years and provides us with an excellent prototype for many generic features of inhomogeneous condensation in dense matter (see, e.g., Refs. \cite{CDW,Adhikari:2016vuu,Andersen:2018osr,zfk,incera,heinz,heinz2,Nickel:2009wj}). Of course, there can be a more favorable form of the inhomogeneous chiral condensate that minimizes thermodynamic potential even more effectively, but using CDW ansatz we can at least conclude that system favours inhomogeneity in some regions. More interesting fact is that recently it was argued in Refs \cite{kojo} that at small temperatures and rather high baryon densities the so-called Quarkyonic phase can be realized in QCD. This phase is described locally by the chiral condensate also in the form of chiral density waves. Furthermore, some realistic parity doublet hadron models, see, e.g., in Ref. \cite{Heinz:2013hza}, predict the CDW phase of dense nuclear matter. 

Duality is another interesting possible property of dense matter, which permits to relate different regions of its phase diagram.
For example, previously in Refs. \cite{ekkz2,ekkz21} it has been found a duality correspondence between CSB and superconductivity 
phenomena on the phase diagram of some low-dimensional NJL-type models extended by baryon and chiral $\mu_5$ chemical potentials.  
It means that if inside some definite 
region of a phase diagram the CSB phase is realized, then in a dually conjugated region the superconductivity must be observed, 
and vice versa. Note that this kind of duality is based on the invariance of the initial NJL-type Lagrangians both in 
two- and three-dimensional spacetimes with respect to some permutation of chemical potentials and coupling constants as well as
discrete  Pauli-Gursey transformations of spinor fields, which connect CSB and superconducting interaction channels 
(see, e.g., in Refs. \cite{ekkz21}). We call this additional invariance of the Lagrangian as its dual symmetry. 
(In contrast, the ordinary Lagrangian symmetry is its invariance with respect to certain field 
transformations only, when external parameters such as chemical potentials etc. remain intact.) Moreover, just due to the 
dual symmetry of the Lagrangian it is also possible to assert that if in the system a particular spatially {\it inhomogeneous} 
phase is realized, then dually transformed phase also exists and it is spatially {\it inhomogeneous}.

In our previous papers \cite{kkz2,kkz,kkzz,kkz3} the phase diagram of dense baryon matter with chiral and isospin imbalances has been 
considered in the framework of the massless NJL model with two quark flavors (see Lagrangian (1) below) both in (3+1)- and 
(1+1)-dimensional spacetime. The model is designed to describe CSB and charged pion condensation phenomena of dense matter. 
Using a homogeneous ansatz for chiral and pion condensates, we have proved in these papers that in the large-$N_c$ limit, 
where $N_c$ is the number of colors, there is a duality correspondence (symmetry) between CSB and charged pion condensation 
(CPC) phenomena on the phase diagram of this effective model. However, at $N_c\ge 3$ this symmetry of the phase diagram (or thermodynamic 
potential) does not correspond to any dual symmetry between CSB and CPC channels of the NJL Lagrangian (1) (at least, we were unable to find such additional 
Lagrangian invariance, typical to some low-dimensional models \cite{ekkz21}, which describes the interaction in CSB and 
superconducting channels).
%similar to that characteristic of Lagrangian, which describes the interaction in CSB and 
%superconducting channels in low dimensions \cite{ekkz21}).  
%However, the Pauli-Gursey type symmetry between CSB and charged pion channels is absent in the model Lagrangian 
%(it agrees with the fact that there is no original Pauli-Gursey symmetry \cite{Stern:1983hg} in $N_c=3$ QCD and, 
%correspondingly, $N_c=3$ NJL model). 
Hence, (i) this kind of duality can be considered only as a dynamical effect, manifested in the symmetry of 
%the leading $1/N_c$-order of the model
the thermodynamic potential (TDP) in the leading $1/N_c$-order of the model with respect to some dual (see below) 
transformations both of order parameters and chemical potentials.  (ii) It is not certain that studies of the phase structure 
of the model in the next orders over $1/N_c$ will give the same result. It is likely that this duality of the phase portrait of 
the NJL model (1) between CSB and CPC phenomena is only approximate. Finally, (iii) the fact that there is a dual symmetry of 
the TDP, calculated in the framework of a homogeneous approach to condensates in the leading $1/N_c$-order, does not at all mean 
that there should be a duality between spatially {\it inhomogeneous} CSB and CPC phases. 
%Hence, a special study is needed to study this last problem. 
\footnote{Earlier, it was found in the paper \cite{he} that  in the 
framework of the (3+1)-dimensional NJL model (1) the spatially inhomogeneous, e.g., CPC phase can be realized in dense 
isotopically asymmetric matter at some chemical potential region. Then, if we had confidence in the existence of a duality 
between nonuniform CSB and CPC phases, we could predict the existence of an inhomogeneous CSB phase in the dually conjugated 
region of chemical potentials.}

The proposed work is devoted to the consideration of this last (iii) problem. In our paper isotopically and chirally asymmetric 
dense baryon matter is investigated, as in Ref. \cite{kkz2}, in the framework of (3+1)-dimensional NJL model. However, 
in contrast to Refs \cite{kkz2}, here we take into account the possibility of spatially inhomogeneous both chiral and charged pion condensates. For the chiral condensate we use the CDW ansatz, while the single-plane-wave LOFF ansatz is used for the charged pion one. Our main result is the conclusion that 
in the leading order of the large-$N_c$ expansion the NJL$_4$ model predicts the duality between these inhomogeneous phases. 
The phase structure itself with inhomogeneous condensates is not studied numerically here and it could be the subject of the future work. But the other method of obtaining information of the phase diagram has been used, namely, the studied duality was employed. We note that earlier some aspects of  the phase diagram of isotopically asymmetric dense baryon matter have been studied in the framework of inhomogeneous condensate approach, and it turns out that it is possible to combine the findings of these studies together and draw full phase diagram. Then, applying just to this phase diagram the dual mapping, we obtain the full phase diagram of chirally asymmetric dense baryon matter. The obtained phase diagram is quite rich and contains various inhomogeneous phases, both   inhomogeneous chiral symmetry breaking phase and inhomogeneous charged pion condensation phase. This example shows that the duality is not just interesting and entertaining mathematical property but a potent instrument with very high predictive capabilities. 

The paper is organized as follows.
First, in Sec. II a (3+1)-dimensional massless NJL model
with two quark flavors ($u$ and $d$ quarks) that  %including
includes four kinds of chemical potentials, $\mu_B,\mu_I,\mu_{I5},\mu_{5}$, is presented. Here spatially inhomogeneous CDW and single-plane-wave LOFF ansatzes respectively for chiral and charged pion condensates are introduced, and the expression for the TDP of the system in the leading $1/N_c$ order is obtained. In Sec. III we show that in the homogeneous approach to condensates the TDP is invariant with respect to three dual symmetries. One of them, ${\cal D}_H$ (\ref{16}), corresponds to the duality of the phase structure of the model between CSB and CPC phenomena. Just this symmetry of the TDP is realized within an inhomogeneous approach. However, two other dual symmetries of the TDP have no analogs in the inhomogeneous approach to the investigation of the ground state of the system. In Sec. IV we show that using only the duality property of dense quark matter, it is possible, basing on the well-known $(\mu_I,\mu)$-QCD phase diagram, to construct the $(\mu_{I5},\mu)$-phase portrait without any numerical studies.   

\section{The model and thermodynamic potential}

We study a phase structure of the two flavored massless (3+1)-dimensional NJL model with several chemical potentials. Its Lagrangian, which is symmetrical under global color $SU(N_c)$ group, has the form
\begin{eqnarray}
&&  L=\bar q\Big [\gamma^\nu\mathrm{i}\partial_\nu
+\frac{\mu_B}{3}\gamma^0+\frac{\mu_I}2 \tau_3\gamma^0+\mu_{5} \gamma^0\gamma^5+\frac{\mu_{I5}}2 \tau_3\gamma^0\gamma^5\Big ]q+ \frac
{G}{N_c}\Big [(\bar qq)^2+(\bar q\mathrm{i}\gamma^5\vec\tau q)^2 \Big
]  \label{1}
\end{eqnarray}
and describes dense quark matter with two massless $u$ and $d$ quarks, i.e. $q$ in (1) is the flavor doublet, $q=(q_u,q_d)^T$, 
where $q_u$ and $q_d$ are four-component Dirac spinors as well as color $N_c$-plets (the summation in (\ref{1}) over flavor, 
color, and spinor indices is implied); $\tau_k$ ($k=1,2,3$) are Pauli matrices. The Lagrangian (1) contains baryon $\mu_B$, 
isospin $\mu_I$, chiral isospin $\mu_{I5}$ and chiral $\mu_5$ chemical potentials. In other words, this model is able to 
describe the properties of quark matter with nonzero baryon $n_B$, isospin $n_I$, chiral isospin $n_{I5}$ and chiral $n_5$ 
densities which are the quantities, thermodynamically conjugated to chemical potentials $\mu_B$, $\mu_I$, $\mu_{I5}$ and $\mu_5$,
respectively. Notice that Lagrangian (1) at low energies describes effectively QCD only at $N_c\ge 3$. At $N_c=2$ QCD is 
additionally symmetric with respect to the so-called Pauli-Gursey symmetry (at zero chemical potentials) and effectively described by another 
NJL-type Lagrangian with diquark interaction channel \cite{weise}.

The quantities $n_B$, $n_I$ and $n_{I5}$ are densities of conserved charges, which correspond to the invariance of Lagrangian (1) with respect to the abelian $U_B(1)$, $U_{I_3}(1)$ and $U_{AI_3}(1)$ groups, where \footnote{\label{f1,1}
Recall for the following that~~
$\exp (\mathrm{i}\alpha\tau_3)=\cos\alpha
+\mathrm{i}\tau_3\sin\alpha$,~~~~
$\exp (\mathrm{i}\alpha\gamma^5\tau_3)=\cos\alpha
+\mathrm{i}\gamma^5\tau_3\sin\alpha$.}
\begin{eqnarray}
U_B(1):~q\to\exp (\mathrm{i}\alpha/3) q;~
U_{I_3}(1):~q\to\exp (\mathrm{i}\alpha\tau_3/2) q;~
U_{AI_3}(1):~q\to\exp (\mathrm{i}
\alpha\gamma^5\tau_3/2) q.
\label{2001}
\end{eqnarray}
So we have from (\ref{2001}) that $n_B=\bar q\gamma^0q/3$, $n_I=\bar q\gamma^0\tau^3 q/2$ and $n_{I5}=\bar q\gamma^0\gamma^5\tau^3 q/2$. We would like also to remark that, in addition to (\ref{2001}), Lagrangian (1) is invariant with respect to the electromagnetic $U_Q(1)$ group,
\begin{eqnarray}
U_Q(1):~q\to\exp (\mathrm{i}Q\alpha) q,
\label{2002}
\end{eqnarray}
where $Q={\rm diag}(2/3,-1/3)$. However, the chiral chemical potential $\mu_5$ does not correspond to a conserved quantity of 
the model (1). It is usually inroduced in order to describe a system on the time scales, when all chirality changing processes 
are finished in the system, and it is in the state of thermodynamical equilibrium with some fixed value of the chiral density 
$n_5$ \cite{andrianov,Cao:2015xja}. 
The ground state expectation values of $n_B$,  $n_I$, $n_{I5}$ and $n_{5}$ can be found by differentiating the thermodynamic 
potential of the system (1) with respect to the corresponding chemical potentials. 

In our previous paper \cite{kkz2} it was shown that if an approach with spatially homogeneous condensates is applied to the 
model (1), then  thermodynamic potential (TDP) of the system obeys some symmetry, which is manifested in the duality between 
CSB and CPC phenomena of dense quark matter (see below). The goal of the present paper is to show that this duality must also 
manifest itself in the approach with inhomogeneous condensates. At the same time, for chiral and charged pion condensates we 
use the CDW and single-plane-wave LOFF ansatzes, respectively.

Let us note that the ansatz used in our present paper includes several possible inhomogeneous condensates but %and 
one may wonder why we do not reproduce the inhomogeneous condensate analyzed by DGR at high baryon densities.
Indeed, in \cite{dgr} finite-density QCD in the large $N_c$ limit has been discussed and it has been noticed that %noticed
at high baryon densities Fermi surface is unstable with respect to the formation of chiral waves with wavenumber 
$2p_F$, where $p_F$ is the Fermi momentum. But they worked in the perturbative regime g$^2 N_c \ll 1$.
In \cite{Shuster:1999tn}, where it was analyzed what happens to the DGR instability at large but
finite $N_c$, the region of the DGR phase was restricted by the line $\mu>3 \Lambda_{QCD} \sim 650-700$ MeV, where $\mu=\mu_B/3$ is 
the quark number chemical potential (maybe in reality, it should be much higher), because below this line the QCD system is 
certainly a strongly-coupled and 
there perturbative calculations cannot be trusted. As a rule, perturbative QCD (without large $N_c$ limit) starts to work 
at very high values, $\mu \gg 1$ GeV \cite{Schafer:1999jg}.
But at that large values of chemical potentials %$\mu$, $\nu$, $\nu_5$ 
the NJL model is not expected to give trustworthy results, as it is an effective low-enegry model for QCD, which is adequate 
only up to the cut-off, which is around 650 MeV. 
And the region of the phase diagram with values of the larger $\mu$ is out of scope of the consideration.
Let us also notice that at large values of $\mu$, even not necessarely as large as 650-700 MeV, the system is probably in the colour superconducting phase. DGR noticed that color superconductivity is suppressed at large $N_c$
due to the fact that the Cooper pair is not a color singlet (the diagram responsible for color superconductivity is non-planar) and only in the large $N_{c}$ limit the DGR phase is predicted.
So summarizing, DGR is an asymtotic phase (at high $\mu$) and cannot be obtained at low values of $\mu$, whereas NJL model desribe only the region of comparatevely low values of $\mu$. 

To find the TDP of the system, we use a semibosonized version of the Lagrangian (\ref{1}), which contains
composite bosonic fields $\sigma (x)$ and $\pi_a (x)$ $(a=1,2,3)$
(in what follows, we use the notations
$\mu\equiv\mu_B/3$, $\nu\equiv\mu_I/2$ and $\nu_{5}\equiv\mu_{I5}/2$):
\begin{eqnarray}
\widetilde L\ds &=&\bar q\Big [\gamma^\rho\mathrm{i}\partial_\rho
+\mu\gamma^0
+ \nu\tau_3\gamma^0+\mu_{5}\gamma^0\gamma^5+\nu_{5}\tau_3\gamma^0\gamma^5-\sigma
-\mathrm{i}\gamma^5\pi_a\tau_a\Big ]q
 -\frac{N_c}{4G}\Big [\sigma\sigma+\pi_a\pi_a\Big ].
\label{2}
\end{eqnarray}
In (\ref{2}) and below the summation over repeated indices is implied. From the auxiliary Lagrangian (\ref{2}) one gets the equations
for the bosonic fields
\begin{eqnarray}
\sigma(x)=-2\frac G{N_c}(\bar qq);~~~\pi_a (x)=-2\frac G{N_c}(\bar q
\mathrm{i}\gamma^5\tau_a q).
\label{200}
\end{eqnarray}
Note that the composite bosonic field $\pi_3 (x)$ can be identified with the physical $\pi^0(x)$-meson field, whereas the physical $\pi^\pm (x)$-meson fields are the following combinations of the composite fields, $\pi^\pm (x)=(\pi_1 (x)\mp i\pi_2 (x))/\sqrt{2}$.   
Obviously, the semibosonized Lagrangian $\widetilde L$ is equivalent to the initial Lagrangian (\ref{1}) when using the equations (\ref{200}).
Furthermore, it is clear from (\ref{2001}) and footnote \ref{f1,1} that the composite bosonic fields (\ref{200}) are transformed under the isospin $U_{I_3}(1)$ and axial isospin $U_{AI_3}(1)$ groups in the following manner:
\begin{eqnarray}
U_{I_3}(1):~&&\sigma\to\sigma;~~\pi_3\to\pi_3;~~\pi_1\to\cos
(\alpha)\pi_1+\sin (\alpha)\pi_2;~~\pi_2\to\cos (\alpha)\pi_2-\sin
(\alpha)\pi_1,\nonumber\\
U_{AI_3}(1):~&&\pi_1\to\pi_1;~~\pi_2\to\pi_2;~~\sigma\to\cos
(\alpha)\sigma+\sin (\alpha)\pi_3;~~\pi_3\to\cos
(\alpha)\pi_3-\sin (\alpha)\sigma.
\label{201}
\end{eqnarray}
Starting from the theory (\ref{2}), one obtains in the leading order of the large $N_c$-expansion (i.e. in the one-fermion loop approximation) the following path integral expression for the effective action ${\cal S}_{\rm {eff}}(\sigma,\pi_a)$ of the bosonic $\sigma (x)$ and $\pi_a (x)$ fields:
$$
\exp(\mathrm{i}{\cal S}_{\rm {eff}}(\sigma,\pi_a))=
  N'\int[d\bar q][dq]\exp\Bigl(\mathrm{i}\int\widetilde L\,d^4x\Bigr),
$$
where
\begin{equation}
{\cal S}_{\rm {eff}}
(\sigma,\pi_a)
=-N_c\int d^4x\left [\frac{\sigma^2+\pi^2_a}{4G}
\right ]+\tilde {\cal S}_{\rm {eff}}
\label{3}
\end{equation}
and $N'$ is a normalization constant. The quark contribution to the effective action, i.e. the term $\tilde {\cal S}_{\rm {eff}}$ in Eq. (\ref{3}), is given by:
\begin{align}%{equation}
\exp(\mathrm{i}\tilde {\cal S}_{\rm {eff}})=N'\int [d\bar
q][dq]\exp\Bigl(\mathrm{i}\int\Big\{\bar q\big
[\gamma^\rho\mathrm{i}\partial_\rho +\mu\gamma^0+\nu\tau_3\gamma^0+
\mu_5\gamma^0\gamma^5+\nu_5\tau_3\gamma^0\gamma^5-\sigma -\mathrm{i}\gamma^5\pi_a\tau_a\big
]q\Big\}d^4 x\Bigr).
 \label{4}
\end{align}%{equation}
The ground state expectation values  $\vev{\sigma(x)}$ and $\vev{\pi_a(x)}$ of the composite bosonic fields are determined by
the saddle point equations,
\begin{eqnarray}
\frac{\delta {\cal S}_{\rm {eff}}}{\delta\sigma (x)}=0,~~~~~
\frac{\delta {\cal S}_{\rm {eff}}}{\delta\pi_a (x)}=0,~~~~~
\label{05}
\end{eqnarray}
where $a=1,2,3$. It is clear from Eq. (\ref{201}) that if $\vev{\sigma(x)}\ne 0$ and/or $\vev{\pi_3(x)}\ne 0$, then the axial isospin $U_{AI_3}(1)$ symmetry of the model is spontaneously broken down, whereas at $\vev{\pi_1(x)}\ne 0$ and/or $\vev{\pi_2(x)}\ne 0$ we have a spontaneous breaking of the isospin $U_{I_3}(1)$ symmetry. Since in the last case the ground state expectation values, or condensates, both of the field $\pi^+(x)$ and of the field $\pi^-(x)$ are nonzero, this phase is usually called the charged pion condensation (CPC) phase. It is easy to see from Eq. (\ref{200}) that the nonzero condensates $\vev{\pi_{1,2}(x)}$ (or $\vev{\pi^\pm(x)}$) are not invariant with respect to the electromagnetic $U_Q(1)$ transformations (\ref{2002}) of the flavor quark doublet. Hence in the CPC phase the electromagnetic $U_Q(1)$ invariance of the model (1) is also broken  spontaneously, so superconductivity is an unavoidable property of the CPC phase.

In vacuum, i.e. in the state corresponding to an empty space with zero particle density and zero values of the chemical potentials $\mu$, $\nu$, $\mu_5$ and $\nu_5$, the quantities $\vev{\sigma(x)}$ and $\vev{\pi_a(x)}$ do not depend on space coordinate $x$. However, in a dense medium, when some of the chemical potentials are nonzero quantities, the ground state expectation values of bosonic fields might have a nontrivial dependence on spatial coordinates. In particular, in this paper we use the following spatially inhomogeneous CDW ansatz for chiral condensate and the single-plane-wave LOFF ansatz for charged pion condensates (for simplicity we suppose that wavevectors of the inhomogeneous condensates are directed along the $x^1$ coordinate axis):
\begin{align}%{eqnarray}
&\vev{\sigma(x)}=M\cos (2kx^{1}),~~~\vev{\pi_3(x)}=M\sin
(2kx^{1}),~~~\nonumber\\
&\vev{\pi_1(x)}=\Delta\cos(2k'x^{1}),~~~
\vev{\pi_2(x)}=\Delta\sin(2k'x^{1}), \label{06}
%~~~~~~~~~06
\end{align}%{eqnarray}
where gaps $M,\Delta$ and wavevectors $k,k'$ are constant dynamical quantities. In fact, they are coordinates of the global minimum point of the TDP $\Omega (M,k,k',\Delta)$. In the leading order of the large-$N_c$ expansion it is defined by the following expression:
\begin{align}
\int d^4x \Omega (M,k,k',\Delta)
=-\frac{1}{N_c}{\cal S}_{\rm {eff}}\{\sigma(x),\pi_a(x)\}\big|_{\sigma
    (x)=\vev{\sigma(x)},\pi_a(x)=\vev{\pi_a(x)}} ,\label{080}
\end{align}
which gives
\begin{align}
\int d^4x\Omega (M,k,k',\Delta)=\int
d^4x\frac{M^2+\Delta^2}{4G}+\frac{\mathrm{i}}{N_c}\ln\left (
\int [d\bar q][dq]\exp\Bigl(\mathrm{i}\int d^4
x\bar q \widetilde{D} q \Bigr)\right ),
\label{08}
\end{align}
where
\begin{eqnarray}%{equation}
\bar q  \widetilde{D} q&=&\bar q\big
[\gamma^\rho\mathrm{i}\partial_\rho
+\mu\gamma^0+\nu\tau_3\gamma^0+\mu_5\gamma^0 \gamma^5+\nu_5\tau_3\gamma^0 \gamma^5-M\exp(2\mathrm{i}\gamma^5\tau_3kx^1)\big ]q\nonumber\\
&-&\Delta\big (\bar q_u\mathrm{i}\gamma^5 q_d\big )\e^{-2\mathrm{i}k'x^1}-\Delta\big (\bar q_d\mathrm{i}\gamma^5 q_u\big )\e^{2\mathrm{i}k'x^1}.\label{09}
\end{eqnarray}%{equation}{equation}
(Remember that in this formula $q$ is indeed a flavor doublet, i.e. $q=(q_u,q_d)^T$.) To proceed, let us introduce in Eqs (\ref{08})-(\ref{09}) the new quark doublets, $\psi$ and $\bar\psi$, by the so-called Weinberg (or chiral) transformation of these fields \cite{kkzz,weinberg}, $\psi=\exp(\mathrm{i}\tau_3k'x^{1}+\mathrm{i}\tau_3\gamma^5kx^{1})q$ and $\bar\psi = \bar q\exp(\mathrm{i}\tau_3\gamma^5kx^{1}-\mathrm{i}\tau_3k'x^{1})$. Since this transformation of quark fields does not change the path integral measure in Eq. (\ref{08})  \footnote{Strictly speaking, performing Weinberg transformation of quark fields in Eq. (\ref{08}), one can obtain in the path integral measure a factor, which however does not depend on the dynamical variables $M$, $\Delta$, $k$, and $k'$. Hence, we ignore this unessential factor in the following calculations. Note that only in the case when there is an interaction between spinor and gauge fields there might appear a nontrivial, i.e. dependent on dynamical variables, path integral measure, generated by Weinberg transformation of spinors. This unobvious fact follows from the investigations by Fujikawa \cite{fujikawa}.}, the expression (\ref{08}) for the TDP is easily transformed to the following one:
\begin{align}%{eqnarray}{eqnarray}
\int d^4x\Omega (M,k,k',\Delta)=\int
d^4x\frac{M^2+\Delta^2}{4G}+\frac{\mathrm{i}}{N_c}\ln\left (
\int [d\bar\psi][d\psi]\exp\Bigl(\mathrm{i}\int d^4
x\bar\psi D\psi \Bigr)\right ),
\label{010}\end{align}
where  instead of the $x-$dependent Dirac operator $\widetilde{D}$ a new $x-$independent operator $D=i\gamma^\mu \partial_{\mu} +\mu\gamma^0 + \nu\tau_3\gamma^0+ \mu_{5}\gamma^0\gamma^5+\nu_{5}\tau_3\gamma^0\gamma^5 +\tau_3\gamma^1 \gamma^5 k+\tau_3\gamma^1 k^{\prime}-M
-\mathrm{i}\gamma^5\Delta\tau_1$ appears. In this case path integral can be evaluated and one get for the TDP (\ref{08}) an expression that reads
\begin{eqnarray}
\Omega (M,\Delta,k,k')~
=\frac{M^2+\Delta^2}{4G}+\mathrm{i}\frac{{\rm
Tr}_{sfx}\ln D}{\int d^4x}=\frac{M^2+\Delta^2}{4G}+\mathrm{i}\int\frac{d^4p}{(2\pi)^4}\ln\Det\overline{D}(p),
\label{07}
\end{eqnarray}
where
\begin{equation}
\overline{D}(p)=\not\!p +\mu\gamma^0
+ \nu\tau_3\gamma^0+\mu_{5}\gamma^0\gamma^5 + \nu_{5}\tau_3\gamma^0\gamma^5 +\tau_3\gamma^1 \gamma^5 k+\tau_3\gamma^1 k^{\prime}-M
-\mathrm{i}\gamma^5\Delta\tau_1\equiv\left
(\begin{array}{cc}
A~, & U\\
V~, & B
\end{array}\right )
\label{500}
\end{equation}
is the momentum space representation of the Dirac operator $D$.
The quantities $A,B,U,V$ in  (\ref{500}) are really the following 4$\times$4 matrices in the spinor space, 
\begin{eqnarray}
&&A=\not\!p +\mu\gamma^0
+ \nu\gamma^0+ \mu_{5}\gamma^0\gamma^5+\nu_{5}\gamma^0\gamma^5-M+\gamma^1 \gamma^5 k+\gamma^1 k^{\prime};\nonumber\\&&B=\not\!p +\mu\gamma^0
- \nu\gamma^0+\mu_{5}\gamma^0\gamma^5- \nu_{5}\gamma^0\gamma^5-M-\gamma^1 \gamma^5 k-\gamma^1 k^{\prime};\nonumber\\&&U=V=-\mathrm{i}\gamma^5\Delta,
\label{80}
\end{eqnarray}
so the quantity $\overline{D}(p)$ from (\ref{500}) is indeed a 8$\times$8 matrix whose determinant in Eq. (\ref{07}) can be calculated on the basis of the following general relations
\begin{eqnarray}
\Det\overline{D}(p)\equiv\det\left
(\begin{array}{cc}
A~, & U\\
V~, & B
\end{array}\right )=\det [-VU+VAV^{-1}B]=\det [\Delta^{2}+ \gamma^5 A \gamma^5 B]=\det
[BA-BUB^{-1}V].
\label{9}
\end{eqnarray}

\section{Duality between CSB and CPC phenomena}
\subsection{Spatially homogeneous approach to condensates}

In this case the wave vectors $k$ and $k'$ in the inhomogeneous ansatzes (\ref{06}) are zero by assumption and, as a result, for the quantity (\ref{9}) one can get the following expression (for details see Ref. \cite{kkz2})
\begin{eqnarray}
\Det\overline{D}(p)=\big (\eta^4-2a_+\eta^2+b_+\eta+c_+\big )\big (\eta^4-2a_-\eta^2+b_-\eta+c_-\big )\equiv P_+(p_0)P_-(p_0),
\label{91}
\end{eqnarray}
where $\eta=p_0+\mu$, $|\vec p|=\sqrt{p_1^2+p_2^2+p_3^2}$ and
\begin{eqnarray}
a_\pm&&=M^2+\Delta^2+(|\vec p|\pm\mu_{5})^2+\nu^2+\nu_{5}^2;~~b_\pm=\pm 8(|\vec p|\pm\mu_{5})\nu\nu_{5};\nonumber\\
c_\pm&&=a_\pm^2-4 \nu ^2
\left(M^2+(|\vec p|\pm\mu_{5})^2\right)-4 \nu_{5}^2 \left(\Delta ^2+(|\vec p|\pm\mu_{5})^2\right)-4\nu^{2} \nu_{5}^2.
\label{10}
\end{eqnarray}
One can notice that in this case the expression (\ref{91}) and hence the TDP (\ref{07}) are invariant with respect to the so-called duality transformation ${\cal D}_H$ of the order parameters and chemical potentials, 
\begin{eqnarray}
{\cal D}_H:~~~~M\longleftrightarrow \Delta,~~\nu\longleftrightarrow\nu_{5}.
 \label{16}
\end{eqnarray}
Other parameters such as $\mu$, $\mu_{5}$ and the coupling constant $G$ do not change under this transformation. It means that if at $\mu,\mu_5,\nu=P,\nu_5=Q$ the global minimum of the TDP lies at the point $(M=M_0,\Delta=\Delta_0)$, then at $\mu,\mu_5,\nu=Q,\nu_5=P$ it is at the point $(M=\Delta_0,\Delta=M_0)$. The results of the paper \cite{kkz2} support this conclusion. In particular, it was shown there that if at some fixed values of chemical potentials, e.g., the homogeneous CSB phase is realized, then in the dually conjugated region of the chemical potentials, i.e. at $\nu\leftrightarrow\nu_{5}$ and unchanged values of $\mu$ and $\mu_5$, the homogeneous CPC phase must be observed in the system and vice versa. 

The duality similar to Eq. (\ref{16}), i.e. the duality between CSB and CPC phenomena, is also observed between phase structures 
of gauge theories with different gauge groups and matter content in the framework of the so-called orbifold equivalence 
formalism in the large-$N_c$ limit \cite{hanada,hanada2}.

In addition to invariance of the TDP with respect to the duality transformation (\ref{16}), in the case of homogeneous approach 
to the condensates there are two other transformations of chemical potentials and order parameters, ${\cal D}_{HM}$ and 
${\cal D}_{H\Delta}$, which we call constrained dual transformations, that leave the TDP unchanged. Indeed, one can check 
(see in Ref. \cite{kkz2}) that under the constraint  $\Delta=0$ and at fixed values of $\mu,\nu$ the TDP (\ref{07}) at $k,k'=0$ is invariant with respect to the permutation $\mu_5\leftrightarrow\nu_{5}$. It is the so-called constrained dual transformation ${\cal D}_{HM}$. Whereas at $M=0$ and at fixed values of $\mu,\nu_5$ the TDP (\ref{07}) at $k,k'=0$ is invariant under the permutation $\mu_5\leftrightarrow\nu$. It is the so-called constrained dual transformation ${\cal D}_{H\Delta}$. The symmetry of the TDP with respect to ${\cal D}_{HM}$ (with respect to ${\cal D}_{H\Delta}$) means that if at some values of the chemical potentials the CSB phase (the CPC phase) is realized, then at $\mu_5\leftrightarrow\nu_{5}$ (at $\mu_5\leftrightarrow\nu$) the same phase will be observed, if dynamically or due to other reasons the charged pion condensate $\Delta$ is equal to zero (the chiral  condensate $M$ is equal to zero) in the system. Hence, in the case of a homogeneous approach to condensates the symmetry of the TDP under the constrained ${\cal D}_{HM}$ and ${\cal D}_{H\Delta}$ dual transformations can also be useful in relating phase structure of the model between dually conjugated regions of the chemical potentials. \label{hom}

Of course, the duality between CSB and CPC phenomena can be broken by expectation values of operators that have not been 
included into the consideration. For example, on the QCD phase diagram there could be more complicated light meson condensations,
such as $\rho$-, $\omega$- meson or kaon, etc. that are not considered in our paper and, in principle, can break the duality of 
the phase structure. Let us 
make a couple of comments %on them
 and show that it is likely that all or a massive region of %the chemical potential space
phase diagram considered
in the paper does not have these condensates and, hence, these condensations can be excluded from the consideration.
It was suggested early on in \cite{VoskresenskySannino} that at sufficiently high $\mu_I$, charged $\rho$-mesons will undergo BEC as pions. %\cite{VoskresenskySannino}. %In terms
%of the holographic model for QCD in Ref. [O. Aharony, K. Peeters, J. Sonnenschein and M. Zamaklar, JHEP 0802 (2008) 071 [arXiv:0709.3948 [hep-th]]] it has been shown that, indeed, $\rho$-mesons
%condense for sufficiently high values of $\mu_I$ ($\mu_I > 1.7m_{\rho}$). This would have had far-reaching consequences for the
%structure of isospin-rich nuclear matter but
But it has been concluded in \cite{Brauner:2016lkh} that $\rho$-meson condensation is possible only at isospin chemical potentials much higher than the  $\rho$-meson mass and, in the context of our consideration,
this is outside of the range that we are interested in (range of validity of NJL model). There is another possibility
such as $\omega$-meson condensation (which is equivalent to inclusion of vector interaction) but, it effectively shifts baryon chemical potential $\mu_B$ and does not spoil the duality. Kaon condensation cannot be considered in two-flavoured NJL model at all. So in our studies we take out of consideration possibility of all the light meson condensation except the pion one.
Also, there is a possibility of colour superconducting phase. If there exist non-zero diquark condensates, then the duality, in general, can be spoiled as well.
But here in our considerations we assumed that it is zero, let us elaborate on that. Color superconductivity arises (in most of the approaches) for chemical potential $\mu$ larger than approximately 350 MeV.
Maybe, it is possible to say that the color superconductivity is not likely realized below 350 MeV and, additionally, it is shown in \cite{Cao:2015xja} that if one includes $\mu_5$ into consideration
the transition to the color superconducting phase shifts to higher values of $\mu$. Nevertheless, all these arguments do not forbid color superconductivity and we should admit that above certain value we neglected a gap
from Cooper pairing just for simplicity.

\subsection{Duality in inhomogeneous case}

Let us now discuss the possibility of the duality between CSB and CPC phenomena when spatially inhomogeneous approach in the form (\ref{06}) to condensates is used, i.e. we suppose that $k\neq 0$, $k'\neq 0$. In this case, using any program of analytic calculations, it is also possible to obtain an exact analytical expression for the quantity (\ref{9}) in the form of the 8-th order polynomial,
\begin{eqnarray}
\Det\overline{D}(p)=a_{8}\eta^8+a_{7}\eta^7+a_{6}\eta^6+a_{5}\eta^5+a_{4}\eta^4+a_{3}\eta^3+a_{2}\eta^2+a_{1}\eta+a_{0},
\label{9100}
\end{eqnarray}
where $\eta =p_0+\mu$, $a_8=1$, $a_7=0$ and %(note that the same values for these coefficients are realized of the polynomial 
%(\ref{91}) are also validequal to zero and one) and
\begin{eqnarray}
&&a_{6}=
-4 \left (k^{\prime 2}+k^2+\Delta ^2+\nu ^2+M^2+\mu_5^2+\nu_{5}^2+|\vec p|^2\right), ~a_{5}=
16 \left (k^{\prime} k \mu_5+\mu_5\nu\nu_5-(k^{\prime} \nu +k
   \nu_5) p_1\right ),
\label{92}
\end{eqnarray}
\begin{eqnarray}
a_{4}&=&
2 \big\{3 k^{\prime 4}+3 k^4+2k^2 k^{\prime 2}+2  k^{\prime 2} \left(\Delta ^2+\nu^2+3 M^2+\mu_5^2+\nu_5^2+p_1^2+3p_2^2+3
   p_3^2\right)+~~~~~~~~~~~~~~~~~\nonumber\\
&&2 k^2 \left(3 \Delta ^2+\nu
   ^2+M^2+\mu_5^2+\nu_5^2+p_1^2+3p_2^2+3
   p_3^2\right)+3 (\Delta ^2+M^2)^2+ 3\nu ^4+3 \nu_5^4\nonumber\\
&&+6( M^2+\Delta^2)|\vec p|^2
   +3 \mu_5^4+3|\vec p|^4+2 \mu_5^2 |\vec p|^2\}+6 \Delta ^2 \nu ^2+6M^2\nu_5^2+\nonumber\\
   &&
  24 k \nu  (\mu_5p_1-k^{\prime} \nu_5)+24
   k^{\prime} \nu_5 (\mu_5p_1-k \nu )+24\mu_5 p_1(
   k^{\prime} \nu_5+ k \nu)+  \nonumber\\
   &&2 \Delta ^2 \nu_5^2+2 M^2\nu ^2+2 (\nu ^2+\nu_5^2)
  \mu_5^2+12\mu_5^2 \left(\Delta ^2+M^2\right)+2 (\nu ^2+\nu_5^2)|\vec  p|^2+\nonumber\\&&
2 \nu ^2 \left(3 \Delta ^2+M^2+\mu_5^2+   \nu_5^2+|\vec p|^2
   \right)+2 \nu_5^2
   \left(\Delta ^2+\nu ^2+3 M^2+\mu_5^2+|\vec p|^2   \right),
\label{93}
\end{eqnarray}
\begin{eqnarray}
a_{3}&=&
32 \big\{k^{\prime 3} (\nu  p_1-k \mu_5)+k^3 (\nu_5 p_1-k^{\prime}\nu_5)+k^{\prime 2} (\nu 
\mu_5 \nu_5-k \nu_5p_1)+k^2 (\nu  \mu_5 \nu_5-k^{\prime} \nu p_1)-\nonumber\\
&&     k^{\prime} k \mu_5 \left(\Delta ^2-\nu^2+M^2+\mu_5^2-\nu_5^2-p_1^2+p_2^2+p_3^2\right)-\nu  \mu_5\nu_5 \left(\Delta ^2+\nu ^2+M^2+\mu_5^2+\nu_5^2-|\vec p|^2\right)\nonumber\\
&&+k^{\prime} \nu  p_1   \left(\Delta ^2+\nu^2+M^2-\mu_5^2-\nu_5^2+|\vec  p|^2\right)+k \nu_5 p_1 \left(\Delta ^2-\nu ^2+M^2-\mu_5^2+\nu_5^2+|\vec p|^2\right)  \big\},...
\label{94}
\end{eqnarray}
We do not give here exact analytical expressions for the coefficients $a_{0,1,2}$ of the polynomial (\ref{9100}), since they are too extensive and take up too much space. Nevertheless, it is possible to check that all the coefficients $a_i$ of the polynomial (\ref{9100}) are invariant with respect to the following duality transformation ${\cal D}_I$ of the chemical potentials, absolute values $\Delta$, $M$ and wavevectors $k$, $k^{\prime} $ of the condensates (\ref{06}) 
\begin{eqnarray}
{\cal D}_I:~~~~M\longleftrightarrow \Delta,~~\nu\longleftrightarrow\nu_{5},~~ k\longleftrightarrow k^{\prime }.
 \label{160}
\end{eqnarray}
(The invariance of $a_{3,4,5,6}$ with respect to the dual transformation (\ref{160}) is directly seen from Eqs (\ref{92})-(\ref{94}).) As a result one can find that the whole TDP $\Omega (M,\Delta,k,k')$ (\ref{07}) of the system is also invariant under the dual transformation ${\cal D}_I$. It means that in the leading order of the large-$N_c$ approximation the duality correspondence between CSB and CPC phenomena, found in \cite{kkz2} for homogeneous case, is valid even in the case if in the system the phases with spatially inhomogeneous condensates are realized. The dual invariance (\ref{160}) of the TDP allows one to perform also a dual mapping of some well-known QCD phase diagrams in order to predict a phase structure of the system under influence of more exotic external conditions, such as chiral asymmetry, etc (see below in the section \ref{V}).

In contrast, in the framework of an inhomogeneous approach to condensates in the form (\ref{06}), we were unable to find analogues of two other dual symmetries, the constrained ${\cal D}_{H\Delta}$ and ${\cal D}_{HM}$ symmetries, which are inherent in the model under consideration within a spatially homogeneous approach to condensates (see in Ref. \cite{kkz2} and/or  the end of the previous subsection \ref{hom}). Indeed, assuming that in Eqs. (\ref{92})-(\ref{94}) $\Delta=0$ and $k'=0$ (or $M=0$ and $k=0$), we see that these coefficients of the polynomial (\ref{9100}) are not invariant with respect to transposition $\nu_5\leftrightarrow\mu_5$ (or $\nu\leftrightarrow\mu_5$). Hence, the analog of the dual symmetry ${\cal D}_{HM}$ (or dual symmetry ${\cal D}_{H\Delta}$) of the TDP in the homogeneous case is not realized in the case, when an inhomogeneous approach to condensates is used. (In fact, in the case of inhomogeneous condensates it is necessary to use more subtle arguments. Indeed, in this approach there arise  usually some spurious (unphysical) terms in the TDP (\ref{07}). So it cannot be considered as a physical TDP of the system. To overcome this difficulty, one should use a more physical regularization procedure or apply to TDP the subtraction procedure using the rule presented, e.g., in Ref. \cite{kkzz} (see there Eqs (47)-(48)). According to it, we can construct the physical TDP $\Omega^{phys} (M,\Delta,k,k')$. It turns out that $\Omega^{phys} $ is invariant under the dual transformation ${\cal D}_I$ (\ref{160}). However, both $\Omega^{phys} (M,\Delta=0,k,k'=0)$ and $\Omega^{phys} (M=0,\Delta,k=0,k')$ are not invariant with respect to the transpositions $\nu_5\leftrightarrow\mu_5$ and $\nu\leftrightarrow\mu_5$, correspondingly, i.e. the constrained dual symmetries are not the properties of the TDP in the case of spatially inhomogeneous approach to condensates.) 

\subsection{Duality and the physical point}

Though the chiral limit is an excellent approximation to QCD, one knows that in reality the current quark masses are nonzero.
Let us now consider briefly the situation with nonzero current quark mass $m_0$ (physical point). The way that one can deal with CDW and/or single plane wave ansatz for charged pion condensate at the physical point is the same as in \cite{Adhikari:2016vuu}. In this case the Lagrangian looks like
\begin{eqnarray}
&&  L=\bar q\Big [\gamma^\nu\mathrm{i}\partial_\nu
+\frac{\mu_B}{3}\gamma^0+\frac{\mu_I}2 \tau_3\gamma^0+\mu_{5} \gamma^0\gamma^5+\frac{\mu_{I5}}2 \tau_3\gamma^0\gamma^5-m_{0}\Big ]q+ \frac
{G}{N_c}\Big [(\bar qq)^2+(\bar q\mathrm{i}\gamma^5\vec\tau q)^2 \Big
]. \label{01} 
\end{eqnarray}
If current quark mass $m_0\ne 0$, then the ansatz (\ref{06}) should be transformed to the following one
\begin{align}%{eqnarray}
&\vev{\sigma(x)}=M\cos (2kx^{1})-m_{0},~~~\vev{\pi_3(x)}=M\sin
(2kx^{1}),~~~\nonumber\\
&\vev{\pi_1(x)}=\Delta\cos(2k'x^{1}),~~~
\vev{\pi_2(x)}=\Delta\sin(2k'x^{1}), \label{006}
%~~~~~~~~~06
\end{align}%{eqnarray}
where gaps $M,\Delta$ and wavevectors $k,k'$ are the same quantities. Using this ansatz in the definition (\ref{080}), one can obtain, instead of Eq. (\ref{010}), the following expression for the TDP $\Omega (M,k,k',\Delta)$ in the leading large-$N_c$ order
\begin{eqnarray}
\int d^4x\Omega (M,\Delta,k,k')~
&=&%\int d^4x\frac{(M\cos (2kx^{1})-m_{0})^2+M^{2}\sin^{2}
%(2kx^{1})+\Delta^2}{4G}+\mathrm{i}{\rm
%Tr}_{sfx}\ln D\nonumber\\&=&
\int d^4x\frac{M^2-2m_{0}M\cos (2kx^{1})+m_{0}^{2}+\Delta^2}{4G}+\mathrm{i}{\rm
Tr}_{sfx}\ln D,\label{key}
\end{eqnarray}
where $D$ is the same $x$-independent Dirac operator as in Eq. (\ref{010}). % we used
%$\vev{\pi_1(x)}^{2}+\vev{\pi_2(x)}^{2}=\Delta^{2}\cos^{2}(2k'x^{1})+\Delta^{2}\sin^{2}(2k'x^{1})=\Delta^{2}$
The averaging over spacetime coordinates in Eq. (\ref{key}) supposes that
$\int d^4x\cos (2kx^{1})=0$ if $k\neq0 $ and $\int d^4x\cos (2kx^{1})=\int d^4x$ if $k=0 $. So, similar to the Ref. \cite{Adhikari:2016vuu}, we get
\begin{eqnarray}
\Omega (M,\Delta,k,k')=
\frac{M^2-2m_{0}M\delta_{k,0}+m_{0}^{2}+\Delta^2}{4G}+\mathrm{i}\int\frac{d^4p}{(2\pi)^4}\ln\Det\overline{D}(p),
\label{070}
\end{eqnarray}
where $\overline{D}(p)$ is the momentum space representation of the Dirac operator $D$ (see in Eq. (\ref{500})). One can see that the only difference compared to the case of zero current quark mass is one term that is proportional to delta symbol. Hence, if the chiral condensate is homogeneous, then it is easy to see that the duality ${\cal D}_{H}$ is no longer the exact symmetry of the TDP, but still one can show that it is a rather good approximation. \footnote{In more detail, the influence of $m_0\ne 0$ on the duality between CSB and CPC phenomena is investigated in \cite{Khunjua:2018jmn} and it is shown there that approximate duality ${\cal D}_{H}$ of the phase diagram is observed in the framework of the NJL$_4$ model, and it is a quite good approximation.}
But if duality between inhomogeneous phases is concerned, the duality is exact even in the case of nonzero current quark mass $m_0$, which is a rather interesting in itself. 

\section{Strength of duality and chirally asymmetric QCD phase diagram }
\label{V}

Recall that chiral asymmetry is a relatively recent phenomenon, so the QCD phase diagram under this condition has not been studied in detail so far. But investigations of the QCD-phase structure at $\mu_5,\mu_{I5}=0$ were performed in the presence of its isospin asymmetry, i.e. at $\mu_I\equiv 2\nu\ne 0$. Let us discuss how it is possible to obtain information about the phase structure of chirally asymmetric dense quark matter (at $\mu_5\ne 0$ and/or $\mu_{I5}\equiv 2\nu_5\ne 0$), applying the so-called duality mapping procedure to the well-known QCD-phase diagram with zero chiral asymmetry (at $\mu_5=0$ and $\nu_5=0$). 

First, let us discuss what one knows about the QCD $(\nu, \mu)$-phase diagram (dense quark matter with isospin asymmetry). 
In the framework of a homogeneous ansatz for condensates the QCD phase diagram has been studied in many approaches and models 
and can be described schematically in the following way (see, e.g., the phase diagram of  Fig. 3 in the paper \cite{he}). If 
the values of isospin chemical potential $\mu_{I}$ is larger than pion mass $m_\pi$, i.e. at $\nu>m_{\pi}/2$,  there is 
homogeneous charged pion condensation (CPC) phase at rather small values of $\mu$. But if $\mu_I$ is less than pion mass, $\nu<m_{\pi}/2$, then homogeneous CSB phase with zero baryon density (it is the so-called vacuum state) is realized for rather small $\mu$. But for rather large $\mu$ the normal quark matter (NQM) phase is arranged (for arbitrary values of $\nu$), in which baryon density is nonzero and chiral symmetry is broken.

This is the sketch of the phase diagram in the physical situation of nonzero current quark mass $m_0$. In the chiral limit 
(zero  current quark mass) it is even simpler for in this case the homogeneous PC phase is realized at any nonzero values of isospin chemical potential provided that the value of the quark number chemical potential $\mu$ is not too large.

Assuming that only inhomogeneous CSB phases are possible in the system, the QCD-phase diagram was obtained within NJL models, e.g., 
in \cite{Nickel:2009wj,Nowakowski:2015ksa,Nowakowski} (similar finding has been obtained in 
\cite{Andersen:2018osr}, where this situation was considered in the chiral limit in the quark-meson model). 
It was found that at rather small values of $\nu$ ($\nu< 60$ MeV) and for $\mu\gtrsim 300$ MeV there can exist a region of 
inhomogeneous CSB (ICSB) phase, namely CDW ansatz has been considered in these papers, and it was shown that though solitonic 
modulations are energetically favored against CDW ansatz, it turned out that these changes of ansatz have only mild influence 
on the phase structure in our model. These calculations were done for simplicity in the chiral limit but probably it is a good 
approximation. The case of non-zero current quark mass in the case of zero isospin asymmetry was considered 
in\cite{Nickel:2009wj} and qualitative picture stays the same. The only changes are that the critical point in the 
$(\mu,T)$-phase diagram shifts towards smaller temperatures and larger quark chemical potentials with increase of current quark 
mass and the region of inhomogeneous phase shrinks because its lower border shifts to the larger values of baryon chemical potential. On the other hand, the possibility of the existence of a spatially inhomogeneous charged pion condensation (ICPC) phase of quark matter has been investigated in the framework of NJL$_4$ model, e.g., in Ref. \cite{he}, where the situation of inhomogeneous CPC and homogeneous CSB phases has been considered. It was shown there that ICPC phase, in which pion condensate exists in a single-plane-wave form, can be realized at a rather high value of $\nu$, $\nu\gtrsim 400$ MeV.  

Let us now try to connect these three situations to get full $(\nu, \mu)$ phase diagram. In \cite{Nowakowski}, where ICSB and 
homogeneous CPC phases have been considered, the CDW was not found in the regions, where CPC phase was considered to be in 
homogeneous case, and CPC and ICPC phases have been found in \cite{%Mu:2010zz
he} (At larger values of $\nu$ the region occupied by ICSB phase decreases but the phase continues to be present at the phase 
diagram up to the values of $\nu=60$ MeV and probably even higher values, there are no plots at larger values in 
\cite{Nowakowski, Nowakowski:2015ksa}. It is not clear how far ICSB phase goes to larger values of $\nu$ but it seems it goes 
over values of $\mu=0.3$ GeV almost to the point $\nu=m_{\pi}/2$). And vise versa in \cite{%Mu:2010zz
he} there has not been found ICPC phase at rather small values of $\nu$, where 
ICSB phase is realized in  \cite{Nowakowski}. And if one assumes that there is no mixed phase with ICSB and ICPC condensates,
then one can attach this figures and sketch the whole $(\nu, \mu)$ phase diagram in inhomogeneous case.
If there is a mixed phase with ICSB and ICPC condensates, then phase diagram can become even more complicated and some regions 
of normal quark matter phase, homogeneous and/or inhomogeneous phases can be exchanged to the mixed phase 
(inhomogeneous) and the inhomogeneous phases can become only larger.
Putting together the results of the study of the QCD phase diagram, performed in the above mentioned papers 
\cite{Nickel:2009wj,Nowakowski:2015ksa,Nowakowski,Andersen:2018osr,he}, etc. both in spatially homogeneous and inhomogeneous 
approaches to order parameters (condensates), as well as in different areas of chemical potential values, and our above 
arguments, one can imagine schematically the following $(\nu, \mu)$-phase portrait of dense quark matter with isotopic 
asymmetry that is depicted in Fig 1. Note that it corresponds to quark matter, in which chiral asymmetry is absent 
($\mu_5=0,\nu_5=0$) and quarks are massive, $m_0\ne 0$. 
\begin{figure}
%----figure 1,2
\includegraphics[width=0.45\textwidth]{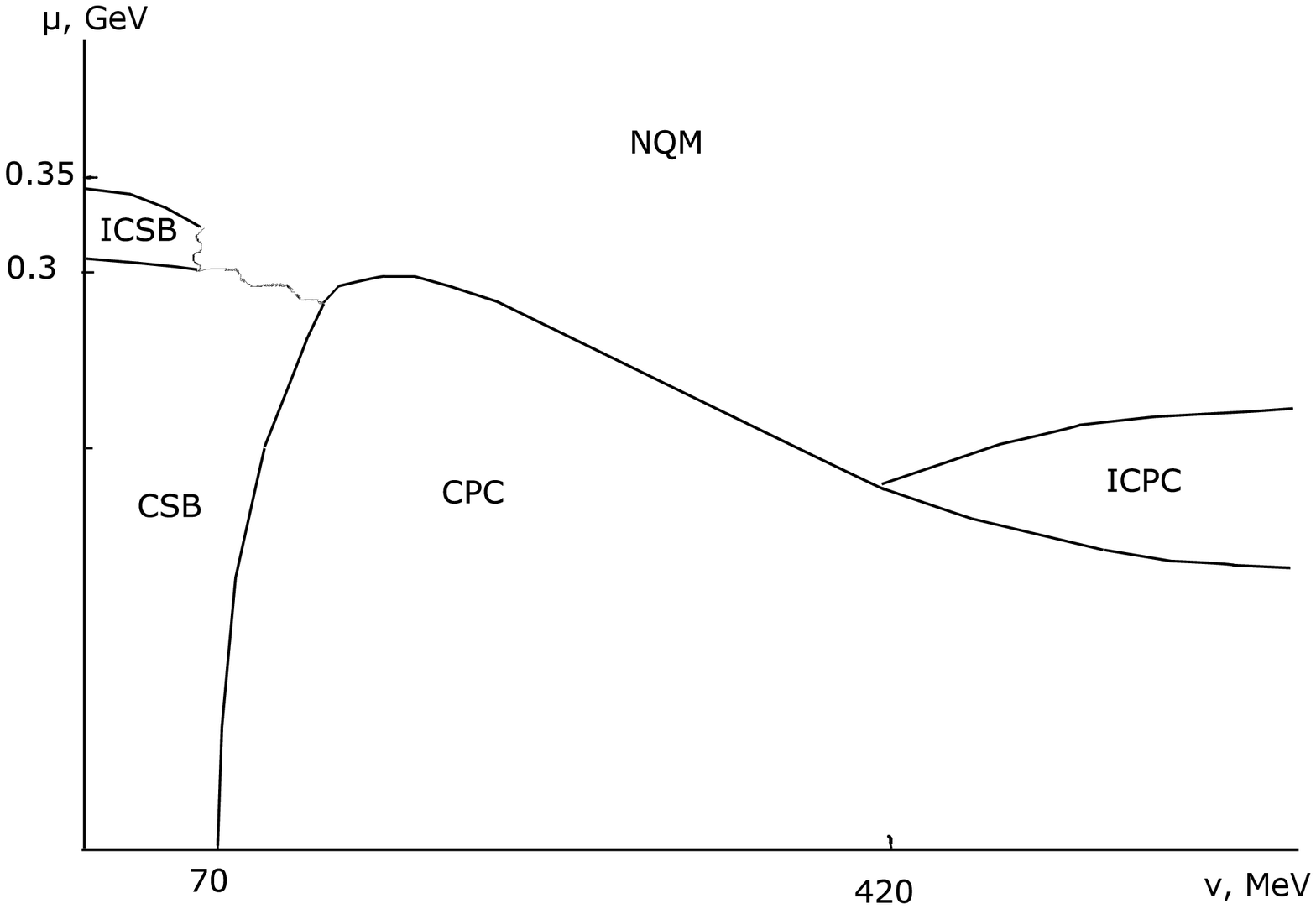}
 \hfill
\includegraphics[width=0.45\textwidth]{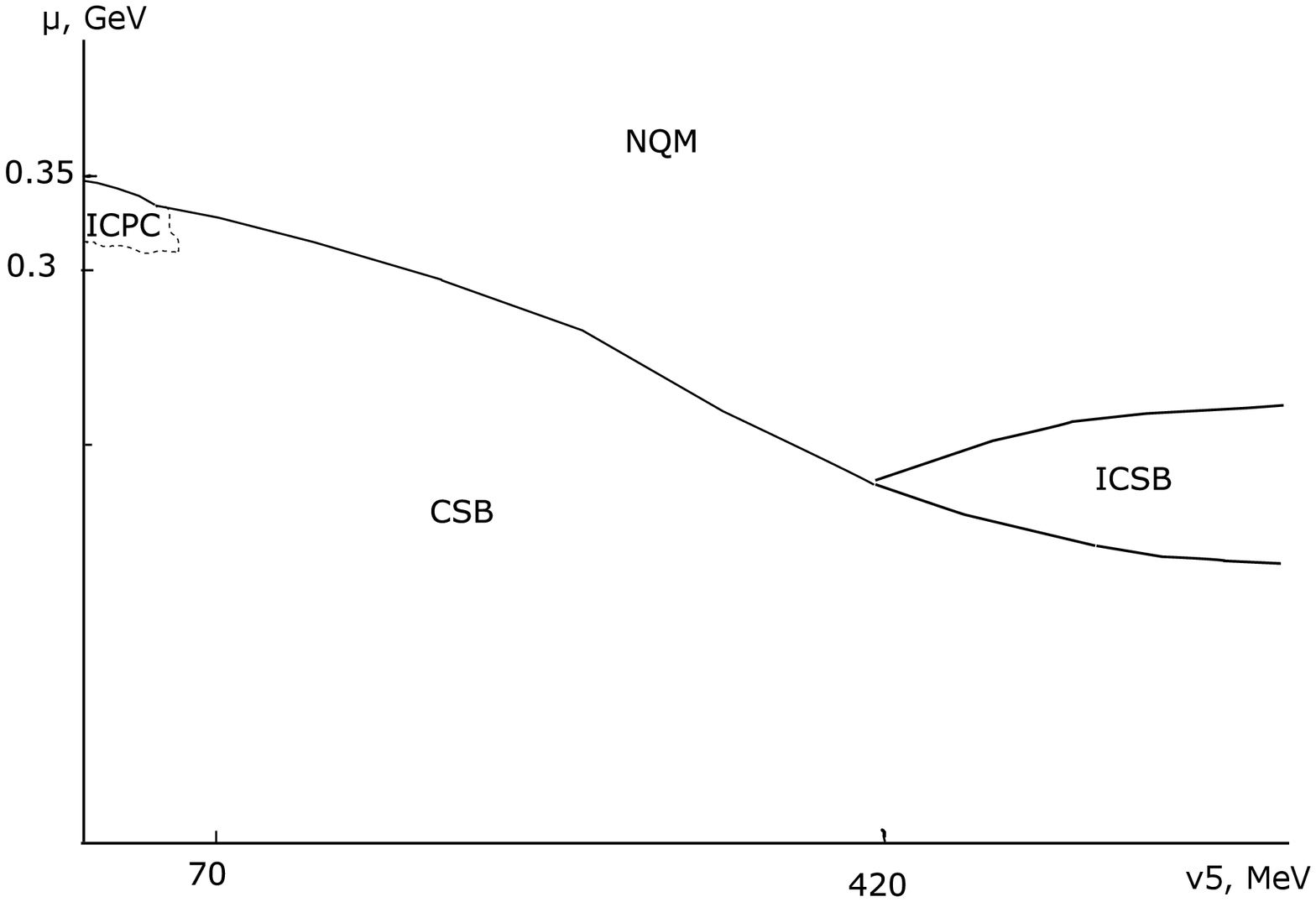}\\
%\label{fig1}
\parbox[t]{0.45\textwidth}{
\caption{Combined  schematic $(\nu,\mu)$-phase diagram at $\mu_{5}=0$ and $\nu_5=0$. Here CPC denotes
the homogeneous charged pion condensation phase, ICPC denotes the inhomogeneous charged pion condensation phase, CSB and ICSB 
denotes homogeneous and inhomogeneous phase with broken chiral symmetry,  NQM is the normal quark matter phase, where charged 
pion condensate is zero and quarks have small masses. }
 }\hfill
\parbox[t]{0.45\textwidth}{
\caption{ $(\nu_{5},\mu)$-phase diagram at $\nu=0$ and $\mu_5=0$. This plot is duality conjugated one to a phase diagram of 
Fig. 1.  All the notations are the same as in Fig. 1. The parts that cannot be drawn  with certainty from the duality only are 
shown with curvy lines. In principle, there could be CPC phase below ICPC one but it is known from other studies that there is no 
homogeneous CPC phase in this region. } }
%\label{fig2}
\end{figure}
Now, let us try to get some information about the phase diagram of dense quark matter when it has chiral asymmetry 
(in the most general case it means that nonzero values of $\mu_5$ and $\nu_5$ should be taken into account in the model). 
Moreover, we suppose that its phase structure is investigated in the framework of a spatially inhomogeneous approach to 
condensates. If $\mu_5\ne 0$, then there is only one way, namely the numerical analysis of the TDP (\ref{07}) or (\ref{070}). 
This task is rather complicated and has not been solved yet. However, if $\mu_5= 0$ but $\nu_5\ne 0$, 
\footnote{As it was discussed in Ref. \cite{Khunjua:2018jmn} (see also Appendix A of the present paper), finite spatial areas 
with nonzero chiral isospin densities, i.e. with $\nu_{5}\ne 0$, might exist inside compact (neutron) stars  due to the chiral 
separation effect. } then, in order to understand how the phase portrait of the model looks, it is not necessary to carry out a numerical study of the TDP (\ref{07}). In this case it is sufficient to perform a dual ${\cal D}_I$ (\ref{160}) mapping  of the phase diagram of Fig. 1 in order to obtain an approximate $(\nu_5,\mu)$-phase portrait at $\nu=0$ and $\mu_5=0$. 

Recall that in the chiral limit and in the leading order of the large-$N_c$ expansion there is an exact dual symmetry between CSB and CPC phenomena predicted by the NJL$_4$ model (1) (see in Refs \cite{kkz3,kkz2}). As a result, there is an (exact) dual correspondence between some phase diagrams, the consequence of which is the possibility to obtain some phase diagrams of the model without cumbersome numerical calculations, simply acting by the dual transformations ${\cal D}_H$ (\ref{16}) or ${\cal D}_I$ (\ref{160}) (see below) on the previously obtained other phase portraits. However, if $m_0\ne 0$, then the duality is only approximate symmetry of the TDP, it is observed in the NJL$_4$ model when some of the chemical potentials are intermediate, i.e., greater than $m_\pi$ (see in Ref. \cite{Khunjua:2018jmn}). And the dual correspondence between phase portraits is absent when chemical potentials are small, i.e. $\lesssim m_\pi$. 

Hence, to get the approximate $(\nu_{5}, \mu)$-phase diagram of the model (\ref{01}) with $m_0\ne 0$ at $\nu=0$ and $\mu_5=0$, 
one just need to take the $(\nu, \mu)$-phase diagram at $\nu_5=0$ and $\mu_5=0$ and make the following transformations of it: 
(i) exchange axis $\nu$ to the axis $\nu_{5}$, (ii) outside the region $\omega=\{(\nu,\mu): 
\nu\lesssim m_\pi, \mu\lesssim 300~ MeV\}$ perform the following renaming of the phases ICSB $\leftrightarrow$ ICPC, CSB 
$\leftrightarrow$ CPC, and NQM phase stays intact here, and (iii) the phase that lies in the region $\omega$ of the 
$(\nu, \mu)$-phase diagram should also be present in the region $\tilde\omega=\{(\nu_5,\mu): \nu_5\lesssim m_\pi, 
\mu\lesssim 300~ MeV\}$ of the $(\nu_5, \mu)$-phase diagram. The obtained phase diagram is shown in Fig. 2 and it is called 
dually ${\cal D}_I$ conjugated to a phase diagram of Fig. 1. Since we suppose that current quark mass $m_0$ is nonzero 
(in this case the dual symmetry between CSB and CPC phenomena is only approximate one \cite{Khunjua:2018jmn}), the dual 
${\cal D}_I$ mapping of the diagram of Fig. 1, i.e. the $(\nu_5, \mu)$-phase diagram of Fig. 2, presents only an approximate 
schematic phase portrait of the model (as far as inhomogeneous phases are concerned it is exact). But, nevertheless, it is 
enough to make several conclusions and predictions about the properties of dense medium with chiral asymmetry.

The obtained phase diagram of Fig. 2 is quite rich and contains an inhomogeneous CSB phase as well as inhomogeneous CPC phase. 
It is clear that (inhomogeneous) charged pion condensation phenomenon can be created in the system at a rather small value of 
$\nu_5$ even at zero value of the chemical potential $\nu$. %(for rather high values of quark number chemical potential $\mu$).
%At small values of $\mu$ chiral isotopic asymmetry in the form of $\nu_5\ne 0$ is able to generate only homogeneous/
%inhomogeneous CSB phase.
Finally, we see that ICPC phase in Fig. 2 is located in the region corresponding to rather high values of $\mu$ and, most likely,
with nonzero baryon densities. Hence, the chiral isotopic asymmetry in the form of $\nu_5\ne 0$ promotes the creation of the 
CPC phenomenon in dense quark matter. Earlier, this effect was established in the framework of a spatially homogeneous approach 
to condensates \cite{kkz3,Khunjua:2018jmn,kkz2} both in the chiral limit and at $m_0\ne 0$, and in the present paper we 
generalize this conclusion to inhomogeneous case, in addition. Let us add a fly in the ointment. One cannot say about the 
presence of the ICPC phase with certainty because the duality for homogeneous condensates is an approximate one and it does not work 
very well in the region of small values of $\nu_{5}$ and $\nu$. And the considerations of \cite{Nowakowski} have not included 
the CPC phase, only the chiral symmetry breaking one was considered. It was possible due to the fact that %because
in the physical situation (at the physical point, non-zero current quark mass) the CPC phase is realized only at the values of 
$\mu_{I}=2\nu$ larger than the value of the pion mass and in that paper there have been considered only smaller values of $\nu$. 
So CPC was not expected in this region and there was no need to compare two minima corresponding to the inhomogeneous CSB phase 
and to the homogeneous CPC phase in the chiral limit. But in the duality conjugated case the homogeneous CSB phase 
(analog of homogeneous CPC phase in the previous case) can be realized at small values of $\nu_{5}$ and in this case to obtain 
the full picture one needs to compare these two phases%(two minima)
. But let us stress that at least, we know for sure that there is a local minimum point 
corresponding to ICPC phase (ICPC phase is boosted by chiral imbalance) and there is another one corresponding to CSB phase. 
To determine which one is the global minimum point, one needs to employ numerical calculations and it cannot be studied in terms of duality 
only. If one assumes that ICPC phase is the global minimum point, then one can see that the chiral isospin chemical potential 
$\nu_{5}$ generates the charged pion condensation phenomenon even better in the inhomogeneous case. And if not, one can say that 
inhomogeneous charged pion condensation phase is favoured over the homogeneous one, but still, chiral symmetry breaking phase is 
a real ground state of the system. Let us also note that, in principle, there could be homogeneous CPC phase at the phase diagram of Fig. 2 below the 
ICPC phase, this fact cannot be figured out just from duality. But from the calculations in the homogeneous case of 
\cite{Khunjua:2018jmn}, it is known that homogeneous CSB phase is a more favourable (its minimum lies lower) than homogeneous 
CPC phase in this region. So, one can conclude that there is no homogeneous CPC phase in this region of the phase diagram,
and the only possibility to have other phases in this region (not CSB phase) is inhomogeneous CPC phase.
One can also see that chiral isospin chemical potential $\nu_{5}$ can lead to the ICSB phase at values of quark chemical 
potential around 0.2 GeV (see Fig. 2). At first glance, one can think that the quark chemical potential is rather small here, 
but due to rather large values of chiral isospin chemical potential, at least a part of this phase can possess non-zero baryon 
density. One can see that at non-zero values of $\nu_{5}$ there seems to be very rich phase diagram featuring as ICSB at rather large 
values of $\nu_{5}$, as probably ICPC phase at small and moderate values of $\nu_{5}$.

This example shows that duality between CSB and CPC phenomena is not just an interesting mathematical artifact, but a powerful tool in scrutinizing the QCD phase diagram. One knows nothing about the phase structure of inhomogeneous condensate in QCD with chiral isospin chemical potential, does not know even whether it is favoured anywhere in the phase diagram at all, and can get the full phase diagram only due to the use of duality.

\section{Summary and conclusions}

In this paper dense quark matter with isospin and chiral imbalance and dualities (symmetries) of its phase diagram are considered in the framework of  (3+1)-dimensional NJL model (1) in the case of spatially inhomogeneous approach to chiral and charged pion condensates.

Earlier, the phase structure of this model has been studied in details in Refs \cite{kkz3,kkz2} in the context of spatially 
homogeneous approach to condensates and in the chiral limit, $m_0=0$. In particular, it was shown there that in the leading 
large-$N_c$ order the TDP of the system is invariant under three different dual transformations, ${\cal D}_{H}$, 
${\cal D}_{H\Delta}$ and ${\cal D}_{HM}$ (see in the present section \ref{hom}). One of them, ${\cal D}_{H}$, is 
realized on the phase portrait of the model as a duality correspondence between CSB and CPC phases. It means that if at some 
fixed values of $\mu$, $\mu_5$, $\mu_{I}=P$ and $\mu_{I5}=Q$, e.g., the CSB (or the CPC) phase is realized in the model, then at the dually conjugated values of the chemical potentials, i.e. at the same values of
$\mu$ and $\mu_5$, but at the permuted values of other chemical potentials, $\mu_{I}=Q$ and $\mu_{I5}=P$, the CPC (or the CSB) 
phase must be arranged. So, it is enough to know the phase structure
of the model at $\mu_{I}<\mu_{I5}$, in order to establish the phase
structure at  $\mu_{I}> \mu_{I5}$. Knowing condensates and other
dynamical and thermodynamical quantities of the system,
e.g., in the CSB phase, one can then obtain the corresponding
quantities in the dually conjugated CPC phase of the model, by simply performing there the duality transformation, $\mu_{I}\leftrightarrow\mu_{I5}$. Two other symmetries of the TDP, ${\cal D}_{HM}$ and ${\cal D}_{H\Delta}$, can also impose some restrictions on the shape of the CSB and CPC phases, respectively (see in \cite{kkz2}). 

Note that similar dualities have been also considered in the framework of universality principle (large-$N_{c}$ orbifold 
equivalence) of phase diagrams in QCD and QCD-like theories in the limit of large $N_{c}$ \cite{hanada,hanada2}. In particular, 
it was noted there that in the chiral limit QCD at $\mu_{I5}\ne 0$ might be equivalent to QCD at $\mu_I\ne 0$, etc (see remarks 
in Sec. 4 of \cite{hanada}). Since the Lagrangian (1) itself does not have a symmetry that would automatically lead to the dual 
symmetries of its phase portrait, an interesting question arise whether the duality of the phase portrait obtained in the 
large-$N_c$ limit is a deep property of the theory described by Lagrangian (1) or it is just an accidental feature. 
\footnote{As a counterexample, we can bring the duality between CSB and superconductivity in some (1 + 1)- and 
(2 + 1)-dimensional theories \cite{ekkz2,ekkz21}. But there the original Lagrangians are invariant with respect to the so-called 
dual symmetry, which includes both the transformation of chemical potentials and coupling constants as well as the Pauli-Gursey 
transformation of spinor fields (that transforms the chiral interaction channel into a superconducting one, and vice versa). 
As a result, a duality between these phenomena appears on the phase portrait.} 
In order to get some hints in this direction, the discussed duality between CSB and CPC phenomena has been established both 
within the homogeneous and inhomogeneous approaches to condensates, but only in the framework of the NJL$_{2}$ model 
\cite{kkz,kkzz} (but in this case the dualities in inhomogeneous and homogeneous approaches are quite similar in terms of 
proving them, and duality in inhomogeneous case is much easier to show). To confirm these results and to ensure that the duality and related phenomena are intrinsic also to (3+1)-dimensional variant of the model (1), in the present paper we study the possibility of dual symmetries of its thermodynamic potential using, in contrast to Ref. \cite{kkz2}, a more extended approach based on the spatially inhomogeneous condensates.

In this paper we have obtained in the leading $1/N_c$ order an exact expression (\ref{07}) of the thermodynamic potential of the model (1), when for chiral and charged pion condensates the CDW and single-plane-wave LOFF ansatzes are used, respectively (see Eq. (\ref{06})). A detailed analysis of the phase structure of the model was not carried out in this case. However, we were able to establish that the TDP (\ref{07}) of the system possesses the dual symmetry ${\cal D}_{I}$ (\ref{160}), which   %, like in the case of a homogeneous approach to condensates (), 
necessarily leads to a duality between CSB and CPC phenomena. Hence, in the chiral limit both in the homogeneous and more extended spatially inhomogeneous approaches to the ground state of the NJL$_4$ system (1), we observe the duality between CSB and CPC phenomena. So, in our opinion, this type of duality is not an artifact of the method of investigation, but the true property of a chirally and isotopically asymmetric dense medium described by the NJL$_4$ Lagrangian (1). 

Moreover, it is known that when nonzero current quark masses is included in the consideration (at the physical point) the 
duality is not exact, though it is a good approximation \cite{Khunjua:2018jmn}. When one considers the duality in the 
inhomogeneous case (between inhomogeneous phases) then the duality is exact even at the physical point (see in section IIIC).

It is necessary to bear in mind that in the model (1) an arbitrary dual invariance of its TDP, calculated in the approach with 
homogeneous condensates, is not automatically transferred to the case of inhomogeneous condensates. Indeed, the duality between CSB and CPC are realized in the model (1) in both approaches, however other dual symmetries, ${\cal D}_{HM}$ and ${\cal D}_{H\Delta}$, of the TDP (\ref{07}) at $k=0$ and $k'=0$ are not observed in the case with inhomogeneous condensates, i.e. at $k\ne 0$ and $k'\ne 0$.

In this paper we have not studied the phase portrait in the framework of (3+1)-dimensional massless NJL model itself, but we have shown that 
even if the phase diagram contains phases with nonzero inhomogeneous condensates, it possesses the property of duality 
(dual symmetry). We demonstrate this fact in terms of TDP in the leading order of the large-$N_c$ approximation.

In the literature, the $(\nu, \mu)$-QCD phase diagram has been studied very intensively and it is understood well in homogeneous case.
The various aspects of the $(\nu, \mu)$-phase diagram with possible inhomogeneous condensates were investigated in \cite{Nowakowski, Nowakowski:2015ksa,he}, etc. It has been shown that it is possible to use these shreds and combine them into one unified picture and draw full $(\nu, \mu)$-phase diagram in inhomogeneous case (see in Fig. 1).
When this interesting thing has been completed, from this assembled phase diagram the phase diagram in a completely different situation has been obtained, namely $(\nu_5, \mu)$-diagram of chirally asymmetric QCD matter (see in Fig. 2). 
It has been shown that at nonzero values of $\nu_{5}$ there is a very rich phase diagram featuring both the ICSB phase at rather 
high values of $\nu_{5}$  and the ICPC phase at small and moderate values of $\nu_{5}$. The phase diagram of Fig. 2 was obtained 
only by using the duality between CSB and CPC phenomena. Moreover, as it was established in Ref. \cite{Khunjua:2018jmn}, 
the existence of the duality between CSB and CPC phenomena is supported by lattice QCD results. These instances show that the 
duality is an inherent property of dense quark matter. So, it is not just entertaining mathematical gaud and interesting, 
but useless mathematical property. In our opinion, it is a potent instrument with very high predictivity power.

It has also been hinted that in inhomogeneous case CPC phase is generated in dense quark matter even by infinitesimally small values of chiral isospin chemical potential $\nu_{5}$. Qualitatively, the same behaviour has been predicted in the framework of (1+1)-dimensional NJL model, this concurrence once more consolidates the confidence that NJL$_{2}$ model can be used as a legit laboratory for the qualitative simulation of specific properties of QCD.

\appendix{}

\section{Generation of nonzero chiral isospin density $n_{I5}$ in dense quark matter}
\label{ApB}

Let us suppose, for simplicity, that dense quark matter consists of two massless $u$ and $d$ quarks, whose chemical potentials, 
$\mu_u=\mu+\nu$ and $\mu_d=\mu-\nu$ (see the notations adopted just before the Eq. (\ref{2})), are positive. Moreover, we suppose also that quarks do not interact, and there is an external magnetic field $\vec B=(0,0,B)$ directed along $z$ axis. In this case in the equilibrium state of quark matter there is a nonzero and nondissipative axial current
\begin{eqnarray}
\vec j_{5f}\equiv\vev{\bar q_f\vec \gamma\gamma^5 q_f}=\frac{Q_f\mu_f\vec B}{2\pi^2}\label{B1}
\end{eqnarray}
for each quark flavor $f=u,d$. In Eq. (\ref{B1})  $Q_f$ is an electric charge of the quark flavor $f$, i.e. $Q_u=2/3$, $Q_d=-1/3$. In this case it is not difficult to conclude from Eq. (\ref{B1}) that axial currents of $u$ and $d$ quarks are opposite in their directions.
Since $\vec j_{5f}=\vev{\bar q_{fR}\vec \gamma q_{fR}}-\vev{\bar q_{fL}\vec \gamma q_{fL}}$, where
\begin{eqnarray}
q_{fR}=\frac{1+\gamma^5}{2}q_f,~~~ q_{fL}=\frac{1-\gamma^5}{2}q_f,\label{B2}
\end{eqnarray}
we see from Eq. (\ref{B1}) that left- and right-handed quarks of each flavor $f=u,d$ moves in opposite directions of 
the $z$ axis. As a result, a spatial separation of quark chiralities for each flavor $f$ occurs. It is the so-called chiral 
separation effect \cite{Metlitski}. In other words, one can say that in the upper half of the three-dimensional space, i.e. 
at $z>0$,  the density, e.g., $n_{uR}\equiv\vev{\bar q_{uR}\gamma^0 q_{uR}}$ of the right-handed $u$ quarks is greater than the 
density $n_{uL}\equiv\vev{\bar q_{uL}\gamma^0 q_{uL}}$ of the left-handed $u$ quarks. Hence, in this case we have at $z>0$ the 
positive values of the chiral density $n_{u5}\equiv n_{uR}-n_{uL}$ for $u$ quarks. (It is evident that at $z<0$ the chiral 
density of $u$ quarks is negative.)

On the contrary, since the axial current $\vec j_{5d}$ of $d$ quarks differs in its direction from the axial current 
$\vec j_{5u}$ of $u$ quarks, one can see that in this case at $z>0$ (at $z<0$) the chiral density $n_{d5}$ of 
of $d$ quarks is negative (positive). Consequently, we have at $z>0$ the positive values of the quantity  
$n_{I5}\equiv n_{u5}-n_{d5}$, which is the ground state expectation value of the chiral isospin density.
%density operator for the chiral isospin charge (it is defined in Eq. (\ref{2004})). 
Whereas at $z<0$ the chiral isospin density is negative.

In summary, we can say that in dense quark medium under the influence of a strong magnetic field (as an example we can mention  
neutral stars), regions with a nonzero chiral isospin density $n_{I5}$ might appear. Therefore physical processes inside these 
regions can be described, e.g., in the framework of the Lagrangians of the form (1), containing chiral isospin 
chemical potential $\mu_{I5}$.

\end{document}